\documentclass[11pt]{article}
\usepackage{amsmath}
\usepackage{amssymb}
\usepackage{graphicx}
\numberwithin{equation}{section} \textwidth=160mm
\begin{document}
\setcounter{page}{0} \thispagestyle{empty}
\begin{flushright}
\end{flushright}
\vspace*{2.0cm}

\begin{center}
{\large\bf Baryogenesis and CP-Violating Domain Walls \\
           in the Background of a Magnetic Field}
\end{center}

\vspace*{1cm}
\renewcommand{\thefootnote}{\fnsymbol{footnote}}

\begin{center}
{L. Campanelli$^{1,2}$\protect\footnote{Electronic address:
              {\tt campanelli@fe.infn.it}},
P. Cea$^{3,4}$\protect\footnote{Electronic address:
              {\tt Paolo.Cea@ba.infn.it}},
G.~L. Fogli$^{3,4}$\protect\footnote{Electronic address:
              {\tt Fogli@ba.infn.it}}, and
L. Tedesco$^{3,4}$\protect\footnote{Electronic address:
              {\tt luigi.tedesco@ba.infn.it}}\\[0.5cm]

$^1${\em Dipartimento di Fisica, Universit\`a di Ferrara, I-44100 Ferrara, Italy}\\[0.2cm]
$^2${\em INFN - Sezione di Ferrara, I-44100 Ferrara, Italy}\\[0.2cm]
$^3${\em Dipartimento di Fisica, Universit\`a di Bari, I-70126 Bari, Italy}\\[0.2cm]
$^4${\em INFN - Sezione di Bari, I-70126 Bari, Italy}}
\end{center}

\vspace*{1.0cm}
\renewcommand{\abstractname}{\normalsize Abstract}


\begin{abstract}
Within the domain wall-mediated electroweak baryogenesis, we study
fermion scattering off a $CP$-violating wall in the background of
an uniform magnetic field. In particular, we calculate the
asymmetry between the reflection coefficients for right-handed and
left-handed chiral fermions, $\Delta R = R_{R \rightarrow L} -
R_{L \rightarrow R}\,$, which is of relevance in non local
baryogenesis mechanisms.
\end{abstract}


\vspace*{1.0cm}
\begin{flushleft} PACS number(s): 12.15.Ji, 11.30.Fs, 98.80.Cq
\end{flushleft}
\vfill
\newpage


\renewcommand{\thesection}{\normalsize{\arabic{section}.}}
\section{\normalsize{Introduction}}
\renewcommand{\thesection}{\arabic{section}}

The origin of the baryon asymmetry in the Universe is a
fundamental question in modern physics. As Sakharov
established~\cite{sakharov} long time ago ( for a review see
Ref.~\cite{trodden} and references therein), elementary particle
physics might account for the production of the baryon asymmetry
in the primordial Universe provided that the interactions meet
some necessary properties. The needed conditions, first identified
by Sakharov, are violation of  $C$ and $CP$ symmetries, violation
of the baryon number $B$,  and departure from thermal equilibrium.
The state of the art of these three conditions in the standard
model is as follows~\cite{trodden}. $B$ violation is obtained
through non perturbative processes that are connected with the
t'Hooft anomaly \cite{hooft} due to the multiple vacuum structure.
Indeed, degenerate vacuum states have different topological
quantum numbers (the so called Chern-Simon number). Particular
configurations called sphalerons \cite{manton} interpolate between
these  vacuum states, thereby leading to processes that violate
baryon number. $C$ and $CP$ symmetries are known to be violated in
the standard model. In particular, $CP$ violation arises from the
quark mass matrix. Finally, the non equilibrium in particle
distributions  is naturally generated at the primordial
electroweak phase transition.
\\
To explain the observed baryon asymmetry one faces with two
problems. First, one needs to understand why in the observable
part of our Universe the baryon number density is many orders
greater than the antibaryon density. Second, one should explain
why the baryon density is much less than the photon density
\cite{kolb}. These aspects are naturally linked to the strength of
$CP$-violation. Indeed, in the standard model $CP$ violation is
extremely weak, so that it is not sufficient to generate the
observed asymmetry. As a matter of fact, it turns out that the
ratio of baryon density $n_B$ to photon density $n_{\gamma}$ comes
out to be of order  $n_B/n_{\gamma} \sim 10^{-20}$ \cite{barr},
that is about ten orders of magnitude smaller than that of recent
measurements from  fluctuations of the cosmic microwave background
by WMAP collaboration~\cite{bennet}:
\begin{equation}
\label{eq1.1} \frac{n_B}{n_{\gamma}} =
\left(6.1^{+0.3}_{-0.2}\right) \times 10^{-10}.
\end{equation}
This clearly indicates that the amount of $CP$ violation in the
standard model is not enough to produce baryon asymmetry at the
observed level. So that we must allow for extra $CP$-violation,
whose origin is probably outside the standard model. In the
present paper we do not address the problem of the origin of $CP$
violation, but we merely parameterize the source of $CP$ violation
by means of an effective complex fermion mass:
\begin{equation}
\label{eq1.2} m(t,\vec{x}) = m_R(t,\vec{x}) + i m_I(t,\vec{x}).
\end{equation}
The second aspect to be considered is the order of the primordial
electroweak phase transition. In the standard electroweak phase
transition the neutral Higgs field is the order parameter which is
expected to undergo a continuum phase transition. Actually, if we
compare the lower bound recently established $M_H > 114.4$ GeV
\cite{hagiwara} with the results of non-perturbative lattice
simulations~\cite{fodor}, we are induced to safely exclude a first
order electroweak phase transition. However, it should be keep in
mind that a first order phase transition can be nevertheless
obtained with an extension of the Higgs sector of the standard
model, or in the minimal supersymmetric standard model.
It is known since long time that, even in a perfectly homogeneous
continuous phase transition, defects will form if the transition
proceeds sufficiently faster than the relaxation time of the order
parameter~\cite{kibble}. In such a non-equilibrium transition, the
low temperature phase starts to form, due to quantum fluctuations,
simultaneously and independently in many parts of the system.
Subsequently, these regions grow together to form the
broken-symmetry phase. When different causally disconnected
regions meet, the order parameter does not generally match and a
domain structure is formed.
In the case in which the phase transition is induced by the  Higgs
sector of the standard model, the defects are domain walls across
which the field flips from one minimum to the other. The defect
density is then related to the domain size and the dynamics of the
domain walls is governed by the surface tension. The existence of
the domain walls, however, is still questionable: it was pointed
out by Zel'dovich, Kobazarev and Okun~\cite{zeldovic} that the
gravitational effects of just one such wall stretched across the
universe would introduce a large anisotropy into the relic
blackbody radiation. For this reason the existence of such walls
was excluded. Quite recently, however, it has been
suggested~\cite{cea1} that the effective surface tension of the
domain walls can be made vanishingly small due to a peculiar
magnetic condensation induced by fermion zero modes localized on
the wall. As a consequence, the domain wall acquires a non zero
magnetic field perpendicular to the wall, and it becomes almost
invisible as far as the gravitational effects are concerned. It
should be stressed, however, that this proposal has been
criticized~\cite{voloshin} (see also Ref.~\cite{cea2}).
Interestingly enough, large-scale magnetic fields have been
observed in galaxies, in galaxy clusters, and there are strong
hints that they exist in superclusters, and in galaxies at high
redshifts. These last astronomical observations support the
conjecture that the magnetic fields have primordial origin
\cite{magnetic}, and then they could have observable effects on
the Cosmic Microwave Background Radiation \cite{Faraday}. Many
mechanisms have been introduced to produce seeds magnetic fields
in the Universe \cite{Dolgov}. A very promising approach relies on
the idea that electroweak phase transition may be the origin of
these seeds \cite{vachaspati}. Indeed, the connection between
continuous phase transitions and primordial magnetic fields has
been studied, for example, in Ref.~\cite{boyanovski}. So that,
primordial magnetic fields could have important consequences on
electroweak baryogenesis~\cite{elmfors}.
\\
The interaction of particles (scalars, Dirac and Majorana
fermions) with domain walls has been the object of various papers
in the literature
\cite{VILENKIN,Iwazaki,Campanelli1,Campanelli2,Cannellos}. In
Ref.~\cite{campanelli} it has been studied the dynamics of
fermions with a spatially varying mass in presence of a
$CP$-violating bubble wall and a uniform magnetic field
perpendicular to the wall. Furthermore the analysis of the
scattering of fermions off a kink and a bubble wall in the case of
a background hypermagnetic field has been done by Ayala et
al.~\cite{ayala,ayala2}.
\\
In this paper we shall consider the dynamics of fermions with a
spatially varying mass in the presence of a $CP$-violating planar
domain wall with an almost uniform magnetic field perpendicular to
the wall. In particular, by using the Dirac equation in presence
of a CP-violating planar domain wall with a constant magnetic
field perpendicular to the wall, we study the fermion scattering
perpendicular to the wall. Neglecting the time dependence of mass
terms, we shall work in the hypothesis that $m(t,\vec x)$ only
depends by the $z$-coordinate perpendicular to the wall, i.e.
$m(t,\vec x)= m(z)$. In the case of domain walls the real part of
the mass term Eq.~(\ref{eq1.2}) goes to $ - m_0$  when $z
\rightarrow - \infty$, while it goes to $m_0$ when $z \rightarrow
+ \infty$, where $m_0$ is fermion mass. We calculate the
reflection coefficients, $R_{R \rightarrow L}$ and $R_{L
\rightarrow R}$, of left-handed and right-handed fermions
respectively. Within the non local baryogenesis
mechanism~\cite{cohen} (or charge transport mechanism) in which
CP-violation and baryon number violation are separated from one
another, the difference $\Delta R = R_{R \rightarrow L} - R_{L
\rightarrow R}$ is relevant for the generation of the baryon
asymmetry~\cite{nelson}.

The plan of the paper is as follows. In Section~2 we set up the
general strategy for solving the Dirac equation in an almost
uniform magnetic field in presence of a CP-violating domain wall.
In Section~2.1, we evaluate the reflection asymmetry specializing
to CP-violating kink domain walls without magnetic field, while in
section~2.2 we consider the effects due to almost uniform magnetic
fields perpendicular to the wall. Finally, we draw our conclusions
in Section~3. Several technical details are relegated in  two
Appendices.


\renewcommand{\thesection}{\normalsize{\arabic{section}.}}
\section{\normalsize{CP-violating Dirac equation in magnetic field}}
\renewcommand{\thesection}{\arabic{section}}

In this Section we analyze the scattering of Dirac fermions off
$CP$-violating domain walls in presence of an electromagnetic
field $A_{\mu}$. We solve the Dirac equation to the first order
with respect to the $CP$-violating term within the so called
distorted-wave Born approximation. After that, we calculate the
reflection coefficients for fermions
and antifermions. \\
In the frame where the wall is at rest, the Dirac equation reads:
\begin{equation}
\label{eq2.1} [\, i / \!\!\!\partial - m(z) P_R - m^{*}(z) P_L - e
\: / \!\!\!\!\!\:\!A \, ] \, \Psi(t,\vec x) = 0,
\end{equation}
where $P_L$ and $P_R$ are the chirality projection operators, $e$
is the electric charge, and $m(z)$ is given by Eq.~(\ref{eq1.2}).
We assume that fermions are coupled to a complex scalar field
through  Yukawa terms. In presence of static domain walls, the
fermion mass term depends only on the distance from the wall. So
that we have $m(t, \vec{x}) = m(z)$, where $z$ is the coordinate
perpendicular to the wall. In order to solve Eq.~(\ref{eq2.1}) we
follow the method of Ref.~\cite{funakubo}. We assume for $\Psi$
the following ansatz:
\begin{eqnarray}
\label{eq2.2}
\Psi(t,z) \!\!& = &\!\! [i / \!\!\!\partial + m(z) P_R + m^{*}(z)
P_L - e \: / \!\!\!\!\!\:\!A \, ] \, e^{-i \sigma E t} \,
\psi_E(z)
\\
\!\!& = &\!\! [i / \!\!\!\partial + m_R - i \, m_I \gamma_5 - e \:
/ \!\!\!\!\!\:\!A \, ] \, e^{-i\sigma E t} \, \psi_E(z),
\end{eqnarray}
with $\sigma = +1 \, (-1)$ for positive (negative) energy
solutions. Moreover, we assume that $A_\mu$ corresponds to a
constant and uniform magnetic field directed along the $z$-axis
with strength $B$. Then, in the Landau gauge we have $A_\mu =
(0,0,-Bx,0)$. Inserting Eq.(\ref{eq2.2}) into Eq.(\ref{eq2.1}) we
obtain:
\begin{equation}
\label{eq2.4}
( E^2 + D_z^2 - |m|^2 + i \gamma^3 D_z m_R + \gamma^3 \gamma_5 D_z
m_I + ieB \gamma^1 \gamma^2 ) \, \psi_E(z) = 0,
\end{equation}
where $D_z = d/dz$. Let us introduce some useful definitions:
\begin{equation}
\label{eq2.5}
 x = az, \;\; \tau = a t, \;\; \epsilon = E/a, \;\;
\xi = m_0/a, \;\; b = eB/a^2,
\end{equation}
where $m_0$ is the fermion mass in the broken phase; $a$ is a
parameter with dimension of mass such that $1/a$ is the
characteristic size of the thickness of the wall. Moreover, we
have:
\begin{eqnarray}
\label{eq2.6}
m_R(z) \!\!& = &\!\! m_0 f(az) = m_0 f(x),
\\
\label{eq2.7}
m_I(z) \!\!& = &\!\! m_0 g(az) = m_0 g(x) \, .
\end{eqnarray}
The function $f(x)$ is the profile of the domain wall, while
$g(x)$ parameterizes the violation of CP.  Using Eq.~(\ref{eq2.6})
and Eq.~(\ref{eq2.7}) we rewrite the Eq.~(\ref{eq2.2}) and
Eq.~(\ref{eq2.4}) respectively as:
\begin{eqnarray}
\label{eq2.8} && \Psi(\tau,x) = (\sigma \epsilon \gamma^0 + i
\gamma^3 D_x + \xi f - i \xi g \gamma_5 + bx \gamma^2) \, e^{-i
\sigma \epsilon \tau} \psi_{\epsilon}(x),
\\
\label{eq2.9} && \left[ \epsilon^2 + D_x^2 - \xi^2 (f^2 + g^2) + i
\xi \gamma^3 f' - \xi \gamma_5 \gamma^3 g'  + ib\gamma^1 \gamma^2
\right] \! \psi_{\epsilon}(x) = 0 \, ,
\end{eqnarray}
where the prime indicates differentiation with respect to $x$. We
shall first solve Eq.~(\ref{eq2.9}), and then use
Eq.~(\ref{eq2.8}) to obtain the complete wave function. For the
time being  we do not need to esplicitate  the functional form of
$f(x)$ and $g(x)$, but we merely assume that
\begin{eqnarray}
\label{eq2.10}
f(x)  & = &
           \begin{cases}
             \,\, +1
             \;\;\;\;\;\;\;\;\;
             \text{for} \; x \rightarrow + \infty
             \\
             \,\, -1
             \;\;\;\;\;\,\;\;\;\;
             \text{for} \; x \rightarrow - \infty
\end{cases}
\end{eqnarray}
and $|g(x)| \ll 1$. The physical meaning of the last relation
comes from the well known fact that $CP$ violation is very small.
It follows that $g(x)$ may be considered as a perturbation. Thus,
we may consistently work our equations  to the first order in
$g(x)$. We begin by splitting the wave function in two terms:
\begin{equation}
\label{eq2.12} \psi_{\epsilon}(x) = \psi^{(0)}(x)  +
\psi^{(1)}(x)\, ,
\end{equation}
where $\psi^{(0)}(x)$ is the solution of the unperturbed equation
obtained from Eq.~(\ref{eq2.9}) setting $g(x)=0$:
\begin{equation}
\label{eq2.13} (\epsilon^2 + D_x^2 - \xi^2 f^2 + i \xi \gamma^3 f'
+ i\, b \, \gamma^1 \gamma^2 ) \, \psi^{(0)}(x) = 0 \, ,
\end{equation}
while $\psi^{(1)}(x)$ is the perturbation, which can be  obtained
by:
\begin{equation}
\label{eq2.14} \psi^{(1)}(x) = \int \! dx' G(x,x') V(x') \,
\psi^{(0)} (x') \, ,
\end{equation}
where $V(x)= - \xi g(x) \gamma_5 \gamma^3$. In Eq.~(\ref{eq2.14})
$G(x,x')$ is the Green's function satisfying:
\begin{equation}
\label{eq2.15} (\epsilon^2 + D_x^2 - \xi^2 f^2 + i \xi f' \gamma^3
+ ib \gamma^1 \gamma^2 )_{\alpha \gamma} G_{\gamma \beta}(x,x') =
- \delta_{\alpha \beta} \delta(x-x') \; ,
\end{equation}
with the same boundary conditions as $\psi^{(0)}(x)$. In order to
find $\Psi(\tau,x)$, we put Eq.~(\ref{eq2.12}) into
Eq.~(\ref{eq2.8}) and obtain:
\begin{equation}
\label{eq2.16} \Psi(\tau,x) \simeq \left[ (\sigma \epsilon
\gamma^0 + i \gamma^3 D_x + \xi f + bx \gamma^2 ) \!
\left(\psi^{(0)} + \psi^{(1)} \right) - i \xi g \gamma_5
\psi^{(0)} \right] \! e^{-i \sigma \epsilon \tau}.
\end{equation}
A standard way to obtain $\psi^{(0)}(x)$ is to expand it in terms
of eigenstates of $\gamma^3$:
\begin{equation}
\label{eq2.17}
\psi^{(0)}(x) = \phi_{\pm}^{(s)}(x) \, u_{\pm}^s \, ,
\end{equation}
with $s=1,2$. The spinors $u_{\pm}^s$ are given by:
\begin{equation}
\label{eq2.18} u_{\pm}^1 = \frac{1}{\sqrt{2}} \! \left(
\begin{array}{c}
1 \\ 0 \\ \pm i \\ 0
\end{array}
\right) \! , \;\;\; u_{\pm}^2 = \frac{1}{\sqrt{2}} \! \left(
\begin{array}{c}
0 \\ 1 \\ 0 \\ \mp i
\end{array}
\right) \! .
\end{equation}
Inserting Eq.~(\ref{eq2.17}) into Eq.~(\ref{eq2.13}) we readily
obtain:
\begin{equation}
\label{eq2.19}
[ \epsilon^2 + D_x^2 - \xi^2 f^2 \mp \xi f' - (-1)^s b ] \,
\phi_{\pm}^{(s)}(x) = 0 \, .
\end{equation}
Let $\phi_{\pm}^{(+\alpha_s)}$ and $\phi_{\pm}^{(-\alpha_s)}$ be
the independent solutions of Eq.~(\ref{eq2.19}), where
\begin{equation}
\label{eq2.20a} \alpha_s  =  i \sqrt{\epsilon^2 - \xi^2 - (-1)^s
b} \; .
\end{equation}
After taking into account the Eq.~(\ref{eq2.10}) we have the
following asymptotic properties for $\phi_{\pm}^{(+\alpha_s)}$ and
$\phi_{\pm}^{(-\alpha_s)}$:
\begin{eqnarray}
\label{eq2.20}
\phi_{\pm}^{(+\alpha_s)}(x)  & \rightarrow &
           \begin{cases}
             e^{+\alpha_s x}
             \;\;\;\;\;\;\;\;\;\;\;\;\;\;\;\;\;\;\;\;\;\;\;\;
             \;\;\;\;\;\;\;\;\;\;\;\;\;\;\;\;\;\;\;\;\;\;\;\;
             \;\;\;\;\;\;\;\;\;\;\;\;
             \text{for} \; x \rightarrow + \infty ,
             \\
             \gamma_{\pm}(\alpha_s,\alpha_s) \, e^{\alpha_s x}+
             \gamma_{\pm}(\alpha_s,-\alpha_s) \, e^{-\alpha_s x}
             \;\;\;\;\;\,\;\;\;\;\;\;\;
             \text{for} \; x \rightarrow - \infty ,
           \end{cases}
\\ \nonumber
\\
\label{eq2.21} \phi_{\pm}^{(-\alpha_s)}(x) & \rightarrow &
           \begin{cases}
             e^{-\alpha_s x}
             \;\;\;\;\;\;\;\;\;\;\;\;\;\;\;\;\;\;\;\;\;\;\;\;
             \;\;\;\;\;\;\;\;\;\;\;\;\;\;\;\;\;\;\;\;\;\;\;\;
             \;\;\;\;\;\;\;\;\;\;\;\;
             \text{for} \; x \rightarrow + \infty ,
             \\
             \gamma_{\pm}(-\alpha_s,\alpha_s) \, e^{\alpha_s x}+
             \gamma_{\pm}(-\alpha_s,-\alpha_s) \, e^{-\alpha_s x} \;\;\;\;\;\;\;
             \text{for} \; x \rightarrow - \infty ,
\end{cases}
\end{eqnarray}
where  $\gamma_{\pm}$ are constants such that
$\gamma_{\pm}(\alpha_s,\alpha_s)^*$ $=\gamma_{\pm}
(-\alpha_s,-\alpha_s)$. Later on we will furnish an explicit
expression for these constants. The general unperturbed solution
can be written as:
\begin{equation}
\label{eq2.23}
\psi^{(0)}(x) = \sum_{s = 1,2} \, [ A_+^{(-\alpha_s)}
\phi^{(-\alpha_s)}_+ (x) + A_+^{(+\alpha_s)} \phi^{(+\alpha_s)}_+
(x) ]\, .
\end{equation}
We are interested in the physical problem where the incident wave
function is coming from $x = -\infty$, it is reflected and
transmitted from the wall at $x=0$, so that at $x = + \infty$
there is only the transmitted wave. In this case it is easy to
find that $A^{(-\alpha_s)}=0$ for $\sigma = +1$, and
$A^{(+\alpha_s)}=0$ for $\sigma = -1$. The next step is to
calculate the Green's function. To this end, following
Ref.~\cite{funakubo} we write:
\begin{equation}
\label{eq2.24}
\Delta_{\alpha \gamma} = (\epsilon^2 + D_x^2 - \xi^2 f^2 + i \xi
f' \gamma^3 + ib \gamma^1 \gamma^2)_{\alpha \gamma} \, ,
\end{equation}
so that Eq.~(\ref{eq2.15}) can be written as:
\begin{equation}
\label{eq2.25}
\Delta_{\alpha\gamma} G_{\gamma \beta} = - \delta_{\alpha \beta}
\delta(x-x') \, .
\end{equation}
Let us introduce the unitary matrix $U$
\begin{equation}
\label{eq2.26}
U=(u^1_+ \;\, u^1_- \;\,  u^2_+ \;\,  u^2_-).
\end{equation}
Taking into account that
\begin{equation}
\label{eq2.28}
U^{-1} \, \gamma^3 \, U  = \left(
\begin{array}{clcr}
i  &       \!&       &  \\
   &   -i  \!&       &  \\
   &       \!&    i  &  \\
   &       \!&       & i
\end{array}
\right) \! ,
\;\;\;\;\;\;
U^{-1} \, \gamma^1 \, \gamma^2 \, U  = \left(
\begin{array}{clcr}
-i &         \!&      &  \\
   &     -i  \!&      &  \\
   &         \!&   i  &  \\
   &         \!&      & i
\end{array}
\right) \! ,
\end{equation}
we can write the matrix
\begin{equation}
\label{eq2.29}
A \equiv U^{-1} \Delta \, U = \left(
\begin{array}{clcr}
\Delta^1_+  &       \!&       &  \\
   &    \Delta^1_-  \!&       &  \\
   &       \!&     \Delta^2_+  &  \\
   &       \!&       &  \Delta^2_-
\end{array}
\right) \! ,
\end{equation}
with
\begin{equation}
\label{eq2.30}
\Delta^{s}_{\pm} = \epsilon^2 + D^2_x - \xi^2 \, f^2(x) \mp \xi \,
f' - (-1)^{s}\,  b \; .
\end{equation}
So that we rewrite Eq.~(\ref{eq2.25}) as
\begin{equation}
\label{eq2.31}
U_{\alpha \gamma} \; A_{\gamma \delta} \;  U_{\delta \rho}^{-1} \;
G_{\rho \beta} (x, x') =  - \delta_{\alpha \beta} \, \delta(x-x').
\end{equation}
Now, writing the Green equation as
\begin{equation}
\label{eq2.32} G(x,x')=U \,  \left(
\begin{array}{clcr}
G^1_+  &       \!&       &  \\
   &    G^1_-  \!&       &  \\
   &       \!&     G^2_+  &  \\
   &       \!&       &  G^2_-
\end{array}
\right) \, U^{-1} \, ,
\end{equation}
Eq.~(\ref{eq2.31}) becomes
\begin{equation}
\label{eq2.33}
U_{\alpha \gamma} \; A_{\gamma \delta} \; U^{-1}_{\delta \rho} \;
U_{\rho k} \; G_{k \sigma} \; U^{-1}_{\sigma \beta} = -
\delta_{\alpha \beta} \, \delta(x-x')\, ,
\end{equation}
from which we simply have
\begin{equation}
\label{eq2.34}
\Delta^s_{\pm} \; G^s_{\pm} = - \delta(x-x')\, .
\end{equation}
In order to determine  the Green's function we consider the
solutions $l^s_{\pm}(x)$ and $m^s_{\pm}(x)$   of the equations
\begin{equation}
\label{eq2.35}
\Delta^s_{\pm} \, l^s_{\pm} (x) =0 \, ,
\;\;\;\;\;\;\;\;\;\;\;\;\;\;\;\;\;\;\;\;\;\;\; \Delta^s_{\pm}
m^s_{\pm}(x)=0
\end{equation}
with the appropriate boundary conditions. Following standard
method \cite{hilbert} we find
\begin{equation}
\label{eq2.36}
G^{(s, \sigma)}_{\pm} (x,x')   =
           \begin{cases}
             - \frac {1} {W(l^s_{\pm}, \,  m^s_{\pm})} \; l_{\pm}(x)
             \; m_{\pm}(x') \;\;\;\;\;\;\;\; x<x'
             \\
             \phantom{.}
             \\
           - \frac {1} {W(l^s_{\pm}, \,  m^s_{\pm})} \; l_{\pm}(x')
             \; m_{\pm}(x) \;\;\;\;\;\;\;\; x'<x
           \end{cases}
\end{equation}
where $W$ is the Wronskian:
\begin{equation}
\label{eq2.37}
W(l^s_{\pm}, \, m^s_{\pm}) = l^s_{\pm} (x) \, \frac {d} {dx} \,
m^s_{\pm}(x) - m^s_{\pm}(x)\,  \frac {d} {dx} \, l^s_{\pm}(x) \; .
\end{equation}
According to our previous discussion, the asymptotic behavior  at
$x \rightarrow + \infty$ is:
\begin{equation}
\label{eq2.38}
m^s_{\pm}(x) = \phi^{(\sigma \alpha_s)}_{\pm}(x) \; ,
\end{equation}
while $l^s_{\pm}$ is  given by:
\begin{equation}
\label{eq2.39}
l^s_{\pm} (x) = \phi^{(-\sigma \alpha_s)}_{\pm} (x) + c^{(s, \,
\sigma)}_{\pm} \, \phi^{(\sigma \, \alpha_s)}_{\pm} (x)
\end{equation}
with $c^{(s, \, \sigma)}_{\pm}$ constants. Since the Wronskian is
constant~\cite{hilbert}, we may evaluate it by considering the
asymptotic forms for $x \rightarrow +\infty$ given by
Eq.(\ref{eq2.20}). It is easy to find:
\begin{equation}
\label{eq2.40}
W(l^s_{\pm}(x), \, m^s_{\pm}(x)) = 2 \, \alpha_s \, \sigma\, ,
\end{equation}
which, indeed, is constant. Putting it all together, the Green's
function becomes:
\begin{equation}
\label{eq2.41}
G_{\pm}^{(s , \sigma)}(x,x') =
                         \begin{cases}
                          - \frac{\sigma}{2 \alpha_s}
                          [\phi_{\pm}^{(-\sigma, \alpha_s)}(x) +
                          c_{\pm}^{(s, \sigma)}
                          \phi_{\pm}^{(+\sigma,
                          \alpha_s)}(x)] \, \phi_{\pm}^{(\sigma,
                          \,
                          \alpha_s)}(x') \;\;\;\;\, \mbox{if} \;\; x < x',
                          \\
                          \phantom{xxxxxx}
                          \\
                          - \frac{\sigma}{2 \alpha_s}
                          [\phi_{\pm}^{(-\sigma, \alpha_s)}(x') +
                          c_{\pm}^{(s, \sigma)}
                          \phi_{\pm}^{(+\sigma,
                          \alpha_s)}(x')] \, \phi_{\pm}^{(\sigma,
                          \,
                          \alpha_s)}(x) \;\;\;\; \mbox{if} \;\; x > x'.
                          \end{cases}
\end{equation}
These results can now be used to calculate $\psi^{(1)}(x)$ from
Eq.~(\ref{eq2.14}). We find:
\begin{eqnarray}
\label{eq2.42}
\psi^{(1)}_s(x)= A^{(+)}_s \frac {\xi (-1)^s u^s_-} {2 \alpha_s}
\!\!\!&  &\!\!\! \!\!\left\{ \phi_-^{(\alpha_s)} (x)
\int^{x}_{-\infty}
[\phi^{(-\alpha_s)}_-(x')-c^{(+)}_-\phi^{(\alpha_s)}_-(x')] \, \,
\phi^{(\alpha_s)}_+(x') \, g'(x') \, dx'
\right. \nonumber
\\
\!\!& + &\!\! \left. [\phi^{(-\alpha_s)}_-(x) + c^{(+)}_-
\phi_-^{(\alpha_s)} (x) ] \, \int^{+\infty}_x  \,
\phi^{(\alpha_s)}_-(x') \, \phi^{(\alpha_s)}_+(x') \, g'(x') dx'
\right\}\ . \nonumber
\\
\phantom {.}
\end{eqnarray}
For later convenience, it is useful to define :
\begin{eqnarray}
\label{eq2.43} I_1^{(s)} \!\!& = &\!\! \int_{-\infty}^{+\infty}
\!\! dx \, g'(x) \, \phi^{(-\alpha_s)}_- (x) \,
\phi^{(+\alpha_s)}_+(x) \, ,
\\
\label{eq2.44} I_2^{(s)} \!\!& = &\!\! \int_{-\infty}^{+\infty}
\!\! dx \, g'(x) \, \phi^{(+\alpha_s)}_- (x) \,
\phi^{(+\alpha_s)}_+(x) \, .
\end{eqnarray}
Moreover we will make use of the following relations:
\begin{eqnarray}
\label{eq2.45}
\gamma^0 u^s_{\pm} \!\!& = &\!\! u^s_{\mp} \, , \nonumber
\\
\gamma^1 \gamma^2 u_{\pm}^s \!\!& = &\!\! i (-1)^s u_{\pm}^s \, ,
\nonumber
\\
\gamma^3 u_{\pm}^s \!\!& = &\!\! \pm i u_{\pm}^s \, ,
\\
\gamma_5 u^s_{\pm} \!\!& = &\!\! \mp i (-1)^s u^s_{\mp} \; , \; s
= 1 , 2 \; . \nonumber
\end{eqnarray}
Actually, we are interested in the limit of $x \rightarrow
+\infty$ for the transmitted wave function, and $x \rightarrow
-\infty$ for the incident and reflected  wave functions. Taking
into account the asymptotic behaviors Eqs.~(\ref{eq2.20}) and
(\ref{eq2.21}) it is easy to determine the relevant wave
functions. For definiteness, we focus on $\sigma = +1$ ($\sigma =
-1$ can be worked out similarly). We have for unperturbed
transmitted, incident and reflected wave functions respectively:
\begin{eqnarray}
\label{eq2.46}
[\psi^{(0)}_s(x)]^{\mbox{{\scriptsize tran}}} \!\!& = &\!\!
A^{(+)}_s u^s_{+} \, e^{\alpha_s x},
\\
\label{eq2.47} [\psi^{(0)}_s(x)]^{\mbox{{\scriptsize inc}}} \!\!&
= &\!\! A^{(+)}_s u^s_{+} \gamma_{+}(\alpha_s, \alpha_s) \,
e^{\alpha_s x},
\\
\label{eq2.48} [\psi^{(0)}_s(x)]^{\mbox{{\scriptsize refl}}} \!\!&
= &\!\! A^{(+)}_s u^s_{+} \gamma_{+}(\alpha_s, - \alpha_s) \,
e^{-\alpha_s x}.
\end{eqnarray}
On the other hand, for the perturbed wave functions, using Eqs.
(\ref{eq2.42}), (\ref{eq2.43}), (\ref{eq2.44}) and the asymptotic
behaviors of $\phi_{\pm}^{(+\alpha_s)}$ and
$\phi_{\pm}^{(-\alpha_s)}$, we find:
\begin{eqnarray}
\label{eq2.49}
 [\psi^{(1)}_s(x)]^{\mbox{{\scriptsize tran}}} \!\!&
= &\!\! A^{(+)}_s \frac{(-1)^s \xi}{2\alpha_s} \, u^s_{-} \!
\left[I^{(s)}_1 + c^{(+)}_{-} I^{(s)}_2 \right] \! e^{\alpha_s x},
\\
\label{eq2.50}
[\psi^{(1)}_s(x)]^{\mbox{{\scriptsize inc}}} \!\!& = &\!\!
A^{(+)}_s \frac{(-1)^s \xi}{2\alpha_s} \, u^s_{-} \, I^{(s)}_2 \!
\left[\gamma_{-}(-\alpha_s, \alpha_s) + c^{(+)}_{-}
\gamma_{-}(\alpha_s, \alpha_s)\right] \! e^{\alpha_s x},
\\
\label{eq2.51}
[\psi^{(1)}_s(x)]^{\mbox{{\scriptsize refl}}} \!\!& = &\!\!
A^{(+)}_s \frac{(-1)^s \xi}{2\alpha_s} \, u^s_{-} \, I^{(s)}_2 \!
\left[\gamma_{-}(-\alpha_s, -\alpha_s) + c^{(+)}_{-}
\gamma_{-}(\alpha_s, -\alpha_s)\right] \! e^{-\alpha_s x}.
\end{eqnarray}
To obtain the asymptotic behavior of the perturbed wave functions
it is enough to insert $\psi^{(0)}(x)$ and $\psi^{(1)}(x)$ into
Eq.~(\ref{eq2.16}). In Appendix~A we evaluate the incident and
reflected wave functions. Using the results in Appendix, we have:
\begin{eqnarray}
\label{eq2.52}
[\Psi_s(x,\tau)]^{\mbox{{\scriptsize inc}}} \!\!& = &\!\!
A^{(+)}_s \gamma_{+}(\alpha_s, \alpha_s) \, e^{-i \epsilon \tau +
\alpha_s x} \nonumber
\\
\!\!& \times &\!\! \left \{ \left[ \epsilon - \frac{(-1)^s \xi
g_{-}}{2} + \frac{(-1)^s \xi (\alpha_s - \xi)}{2\alpha_s} \,
\frac{\gamma_{-}(-\alpha_s, \alpha_s)}{\gamma_{+}(\alpha_s,
\alpha_s)} \, I^{(s)}_2 \right] \right. \! u^s_{-} \nonumber
\\
\!\!& + &\!\! \left[ \left. -\alpha_s - \xi + \frac{(-1)^s \xi
\epsilon g_{-}}{2(\alpha_s - \xi)} + \frac{(-1)^s \xi
\epsilon}{2\alpha_s} \, \frac{\gamma_{-}(-\alpha_s,
\alpha_s)}{\gamma_{+}(\alpha_s, \alpha_s)} \, I^{(s)}_2 \right ]
\! u^s_+ \right \} ,
\end{eqnarray}
\begin{eqnarray}
\label{eq2.53}
[\Psi_s(x,\tau)]^{\mbox{{\scriptsize refl}}} \!\!& = &\!\!
A^{(+)}_s \gamma_{+}(\alpha_s, -\alpha_s) \, e^{-i \epsilon \tau -
\alpha_s x} \nonumber
\\
\!\!& \times &\!\! \left \{ \left[ \epsilon - \frac{(-1)^s \xi
g_{-}}{2} - \frac{(-1)^s \xi (\alpha_s + \xi)}{2\alpha_s} \,
\frac{\gamma_{-}(-\alpha_s, -\alpha_s)}{\gamma_{+}(\alpha_s,
-\alpha_s)} \, I^{(s)}_2 \right ] \right. \! u^s_{-} \nonumber
\\
\!\!& + &\!\! \left[ \left. \alpha_s - \xi - \frac{(-1)^s \xi
\epsilon g_{-}}{2(\alpha_s + \xi)} + \frac{(-1)^s \xi
\epsilon}{2\alpha_s} \, \frac{\gamma_{-}(-\alpha_s,
-\alpha_s)}{\gamma_{+}(\alpha_s, -\alpha_s)} \, I^{(s)}_2 \right ]
\! u^s_{+} \right \},
\end{eqnarray}
where $g_{-} \equiv \lim_{x \rightarrow -\infty} g(x)$. The next
step consists in the calculation of the vectorial and axial
currents $j^{\mu}_V = \bar{\Psi} \gamma^{\mu} \Psi$  and
$j^{\mu}_A = \bar{\Psi} \gamma^{\mu} \gamma_5 \Psi$.


\renewcommand{\thesection}{\normalsize{\arabic{section}}}
\subsection{\normalsize{Reflection asymmetry:  $B=0$}}
\renewcommand{\thesection}{\arabic{section}}

Before considering the general case of a uniform magnetic field,
for completeness we consider the currents for $B=0$. We define:
\begin{eqnarray}
\label{eq2.55} \delta^{\mbox{{\scriptsize inc}}} \!\!& = &\!\!
\frac{\xi}{\epsilon} \, \text{Re} \! \left[ \frac{\alpha -
\xi}{\alpha} \, \frac{\gamma_{-}(-\alpha,
\alpha)}{\gamma_+(\alpha, \alpha)} \, I_2 \right] \! ,
\\
\label{eq2.56} \delta^{\mbox{{\scriptsize refl}}} \!\!& = &\!\!
\frac{\xi}{\epsilon} \, \text{Re} \! \left[ \frac{-\alpha -
\xi}{\alpha} \, \frac{\gamma_{-}(-\alpha,
-\alpha)}{\gamma_+(\alpha, -\alpha)} \, I_2 \right] \! ,
\end{eqnarray}
where $\alpha = i \sqrt{\epsilon^2 - \xi^2}$, and $\text{Re}[x]$
is the real part of $x$. Note that in this case the currents,
$\alpha$, and the constants $\gamma_{\pm}$ do not depend on the
spin $s$. So that also $I_2$, defined by Eq.~(\ref{eq2.44}), does
not depend on the spin. The transmitted, incident and reflected
axial currents are given by:
\begin{eqnarray}
\label{eq2.57} && (j^3_A)^{\mbox{{\scriptsize tran}}}_{B=0} = 2
\epsilon^2 \sum_s \, (-1)^{s+1} |A^{(+)}_s|^2,
\\
\label{eq2.58} && (j^3_A)^{\mbox{{\scriptsize inc}}}_{B=0} = 2
\epsilon^2 |\gamma_{+}(\alpha, \alpha)|^2 \sum_s \, (-1)^{s+1}
|A^{(+)}_s|^2 \, [1 - (-1)^{s+1} \delta^{\mbox{{\scriptsize
inc}}}],
\\
\label{eq2.59} && (j^3_A)^{\mbox{{\scriptsize refl}}}_{B=0} = 2
\epsilon^2 |\gamma_{+}(\alpha, -\alpha)|^2 \sum_s \, (-1)^{s+1}
|A^{(+)}_s|^2 \, [1 - (-1)^{s+1} \delta^{\mbox{{\scriptsize
refl}}}].
\end{eqnarray}
Moreover, analogously the vectorial currents turn out to be:
\begin{eqnarray}
\label{eq2.60} && (j^3_V)^{\mbox{{\scriptsize tran}}}_{B=0} = 2
\epsilon \sqrt{\epsilon^2 - \xi^2} \sum_s |A^{(+)}_s|^2,
\\
\label{eq2.61} && (j^3_V)^{\mbox{{\scriptsize inc}}}_{B=0} = - 2i
\alpha \epsilon |\gamma_{+}(\alpha, \alpha)|^2 \sum_s
|A^{(+)}_s|^2 \, [1 - (-1)^{s+1} \delta^{\mbox{{\scriptsize
inc}}}],
\\
\label{eq2.62} && (j^3_V)^{\mbox{{\scriptsize refl}}}_{B=0} =  2i
\alpha \epsilon |\gamma_{+}(\alpha, -\alpha)|^2 \sum_s
|A^{(+)}_s|^2 \, [1 - (-1)^{s+1} \delta^{\mbox{{\scriptsize
refl}}}].
\end{eqnarray}
Then, we have for the reflection coefficient:
\begin{equation}
\label{eq2.63} R^{(0)} = - \frac{(j^3_V)^{\mbox{{\scriptsize
refl}}}_{B=0}}{(j^3_V)^{\mbox{{\scriptsize inc}}}_{B=0}} = \left|
\frac {\gamma_+(\alpha,-\alpha)} {\gamma_+(\alpha,\alpha)}
\right|^2 .
\end{equation}
Our goal is to calculate
\begin{equation}
\label{eq2.64} \Delta R = R_{R \rightarrow L} - R_{L \rightarrow
R} \, ,
\end{equation}
where we have indicate with
\begin{eqnarray}
\label{eq2.65} R_{R \rightarrow L} \!\!& = &\!\! -
\frac{(j^3_{L})^{\mbox{{\scriptsize
refl}}}}{(j^3_{R})^{\mbox{{\scriptsize inc}}}}
 = \frac{{(j^3_{A})^{\mbox{{\scriptsize refl}}}} -
 {(j^3_{V})^{\mbox{{\scriptsize refl}}}}}{{(j^3_{V})^{\mbox{{\scriptsize inc}}}} +
 {(j^3_{A})^{\mbox{{\scriptsize inc}}}}} \: ,
\\ \nonumber
\\
\label{eq2.66}
R_{L \rightarrow R} \!\!& = &\!\! - \frac
{(j^3_{R})^{\mbox{{\scriptsize
refl}}}}{(j^3_{L})^{\mbox{{\scriptsize inc}}}} =
\frac{{(j^3_{A})^{\mbox{{\scriptsize refl}}}} \, + \,
{(j^3_{V})^{\mbox{{\scriptsize
refl}}}}}{{(j^3_{A})^{\mbox{{\scriptsize inc}}}} \, - \,
{(j^3_{V})^{\mbox{{\scriptsize inc}}}}} \:
\end{eqnarray}
the reflection coefficients for right-handed and left-handed
chiral fermions, respectively.
\\
A standard calculation gives:
\begin{equation}
\label{eq2.67} \Delta R = \frac{2\epsilon R^{(0)}}{|\alpha|}
(\delta^{\mbox{{\scriptsize inc}}} - \delta^{\mbox{{\scriptsize
refl}}}).
\end{equation}
It is interesting to consider domain walls with kink wall profile:
\begin{equation}
\label{kink}
f(x) \; = \; \tanh x \; \; .
\end{equation}
The scattering of fermions off  kink domain walls has been
discussed for the first time in Ref.~\cite{voloshin2}. In this
case, one may explicitly solve Eq.~(\ref{eq2.19}). Indeed, it is
easy to find that the solution of Eq.~(\ref{eq2.19}) is:
\begin{equation}
\label{eq2.80} \phi_{\pm}^{(+ \alpha)}(y)  =  y^{-\alpha/2} \,
(1-y)^{\alpha/2} \, _2 F_1 \left[ \,  1 \mp \xi/2, \, \pm \xi/2,
\, 1 - \alpha\, ; \,  y \right],
\end{equation}
where $y =(1- \tanh x)/2$ and $_2 F_1 [a,b,c;z]$ is the
hypergeometric function. As a consequence, we may write down the
explicit expressions for $\gamma_{\pm}(\alpha, \alpha)$ and
$R^{(0)}$. We find:
\begin{equation}
\label{eq2.81} \gamma_{\pm}(\alpha, \alpha) = \frac {
\Gamma(-\alpha + 1) \, \Gamma(-\alpha)} {\Gamma[(-2 \alpha \pm
\xi)/2] \, \Gamma[(-2 \alpha \mp \xi)/2 +1]} \, ,
\end{equation}
\begin{equation}
\label{eq2.82} R^{(0)} = \frac{\cos(\pi \xi) -1}{\cos(\pi \xi) -
\cosh(2 \pi |\alpha|)} \, ,
\end{equation}
which indeed agree with Ref.~\cite{voloshin2}. As concern the
function $g(x)$, we shall follow the method of
Ref.~\cite{campanelli,funakubo}, where one finds two functional
forms for $g(x)$, namely  $g(x) = \Delta \theta f^2(x)$ and $g(x)
= \Delta \theta f'(x)$. The parameter $\Delta \theta$ measures the
strength of $CP$-violation, so that we assume $|\Delta \theta| \ll
1$.
\\
In Fig.~1 we display $\Delta R / \Delta \theta$ as a function of
the normalized energy $\epsilon^* = \epsilon a = E/m_0$, for
different values of the thickness of the wall $a^* = a/m_0$, in
the cases $g = \Delta \theta f'$ and $g = \Delta \theta f^2$. We
observe that $\Delta R$ may be positive or negative depending on
the functional form of $g$, and that at fixed thickness of the
wall (proportional to 1/$a^*$), the absolute value of $\Delta R$
goes to zero when the energy of the incident particles approaches
to infinity.


\vspace{1.0truecm}

\begin{figure}[h!]
\begin{center}
\includegraphics[clip,width=0.48\textwidth]{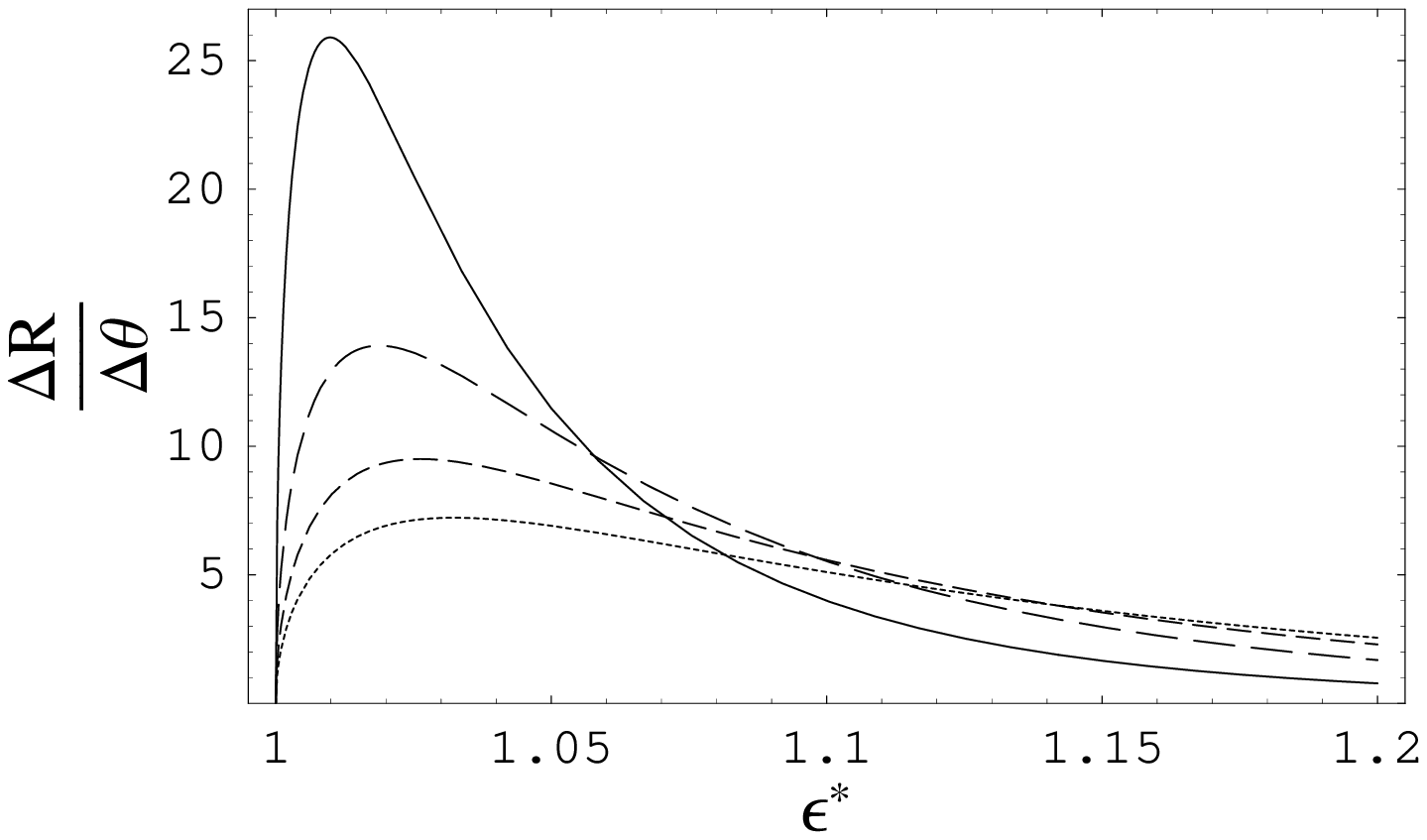}
\hskip 0.5truecm
\includegraphics[clip,width=0.48\textwidth]{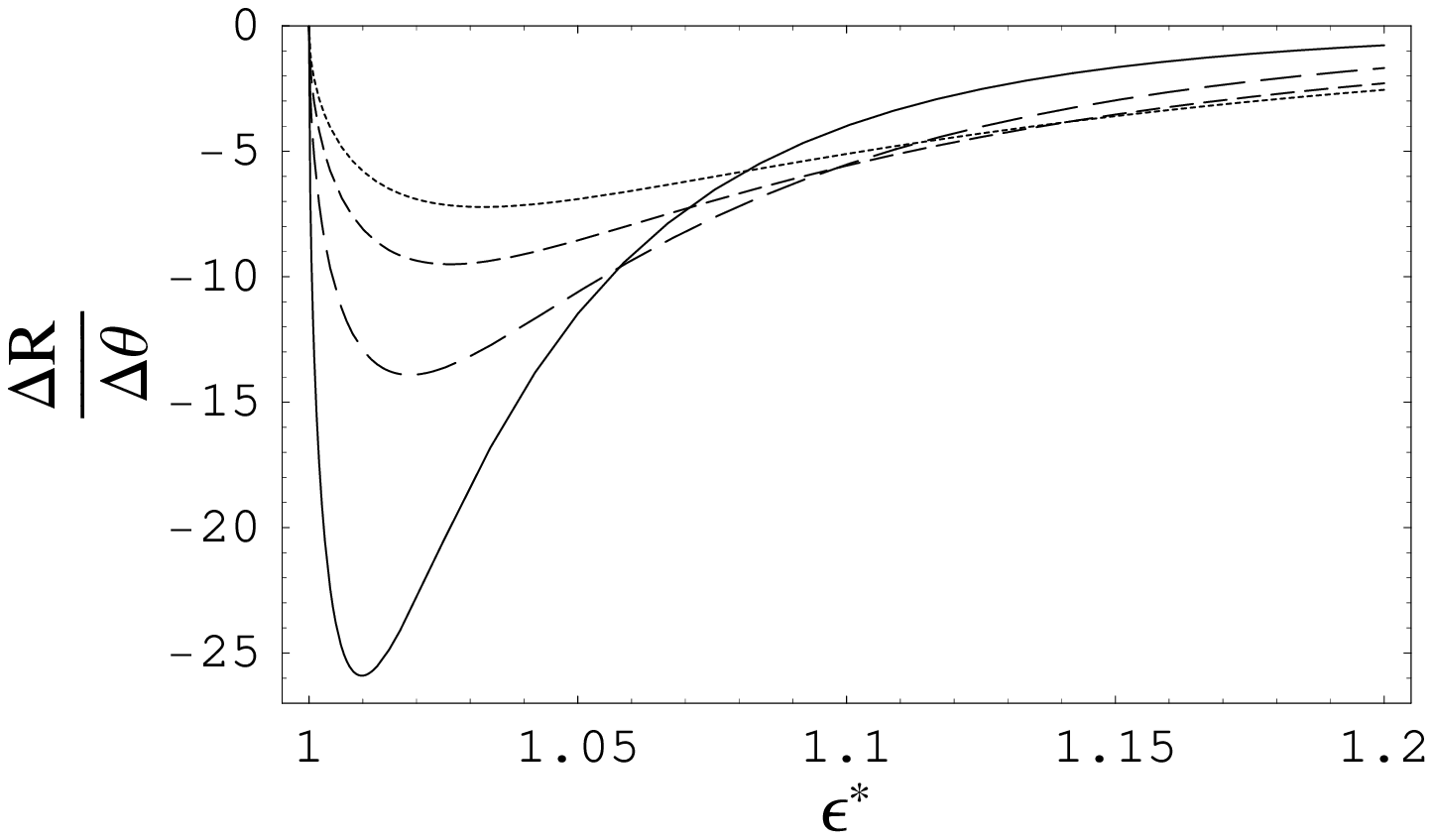}
\caption{Case B=0. $\Delta R / \Delta \theta$ versus $\epsilon^*$
in the cases $g = \Delta \theta f'$ (left panel) and $g = \Delta
\theta f^2$ (right panel) for four different values of $a^*$.
Solid line: $a^*=0.8$; long-dashed line: $a^*=1$; short-dashed
line: $a^*=1.2$; dotted line: $a^*=1.4$.}
\end{center}
\end{figure}


\renewcommand{\thesection}{\normalsize{\arabic{section}}}
\subsection{\normalsize{Reflection asymmetry: $B\neq0$}}
\renewcommand{\thesection}{\arabic{section}}

Let us consider the most physically relevant case when $B \neq 0$.
In Appendix~B we show that we may write:
\begin{eqnarray}
\label{eq2.68a} (j^3_{V,A,s})^{\mbox{{\scriptsize inc}}} \!\!& =
&\!\! (j^3_{V,A,s})_{(0)}^{\mbox{{\scriptsize inc}}} \, (1 +
\delta_{V,A,s}^{\mbox{{\scriptsize inc}}})
\\
\label{eq2.69a} (j^3_{V,A,s})^{\mbox{{\scriptsize refl}}} \!\!& =
&\!\! (j^3_{V,A,s})_{(0)}^{\mbox{{\scriptsize refl}}} \, (1 +
\delta_{V,A,s}^{\mbox{{\scriptsize refl}}})
\end{eqnarray}
where
\begin{eqnarray}
\label{eq2.70a} (j^3_{V,s})_{(0)}^{\mbox{{\scriptsize inc}}} \!\!&
= -&\!\! 2 \,
 i \, \epsilon \, \alpha_s \, |A^{(+)}_s|^2 \, |\gamma_{+}(\alpha_s,
\alpha_s)|^2
\\
\label{eq2.71a} (j^3_{V,s})_{(0)}^{\mbox{{\scriptsize refl}}}
\!\!& = &\!\! 2 \,
 i \, \epsilon \, \alpha_s \, |A^{(+)}_s|^2 \, |\gamma_{+}(\alpha_s,
-\alpha_s)|^2
\\
\label{eq2.72a} (j^3_{A,s})_{(0)}^{\mbox{{\scriptsize inc}}} \!\!&
= &\!\! (-1)^s \, |A^{(+)}_s|^2 \, |\gamma_{+}(\alpha_s,
\alpha_s)|^2 \, [2 \, \epsilon^2 - (-1)^s \, b]
\\
\label{eq2.73a} (j^3_{A,s})_{(0)}^{\mbox{{\scriptsize refl}}}
\!\!& = &\!\! (-1)^s \, |A^{(+)}_s|^2 \, |\gamma_{+}(\alpha_s,
-\alpha_s)|^2 \, [2 \, \epsilon^2 - (-1)^s \, b].
\end{eqnarray}

\vspace{1.0truecm}


\vspace{1.0truecm}

\begin{figure}[ht]
\begin{center}
\includegraphics[clip,width=0.48\textwidth]{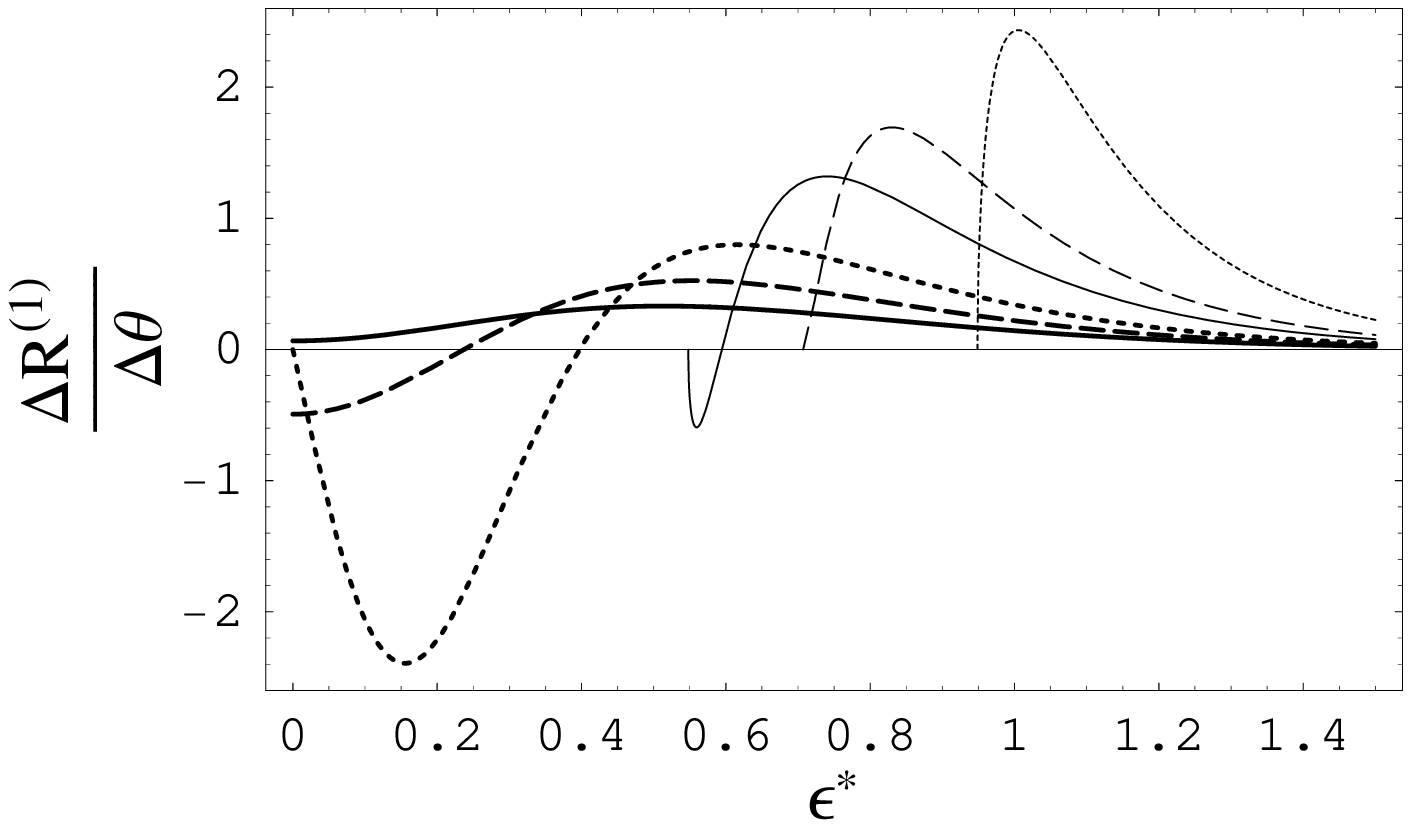}
\hskip 0.5truecm
\includegraphics[clip,width=0.48\textwidth]{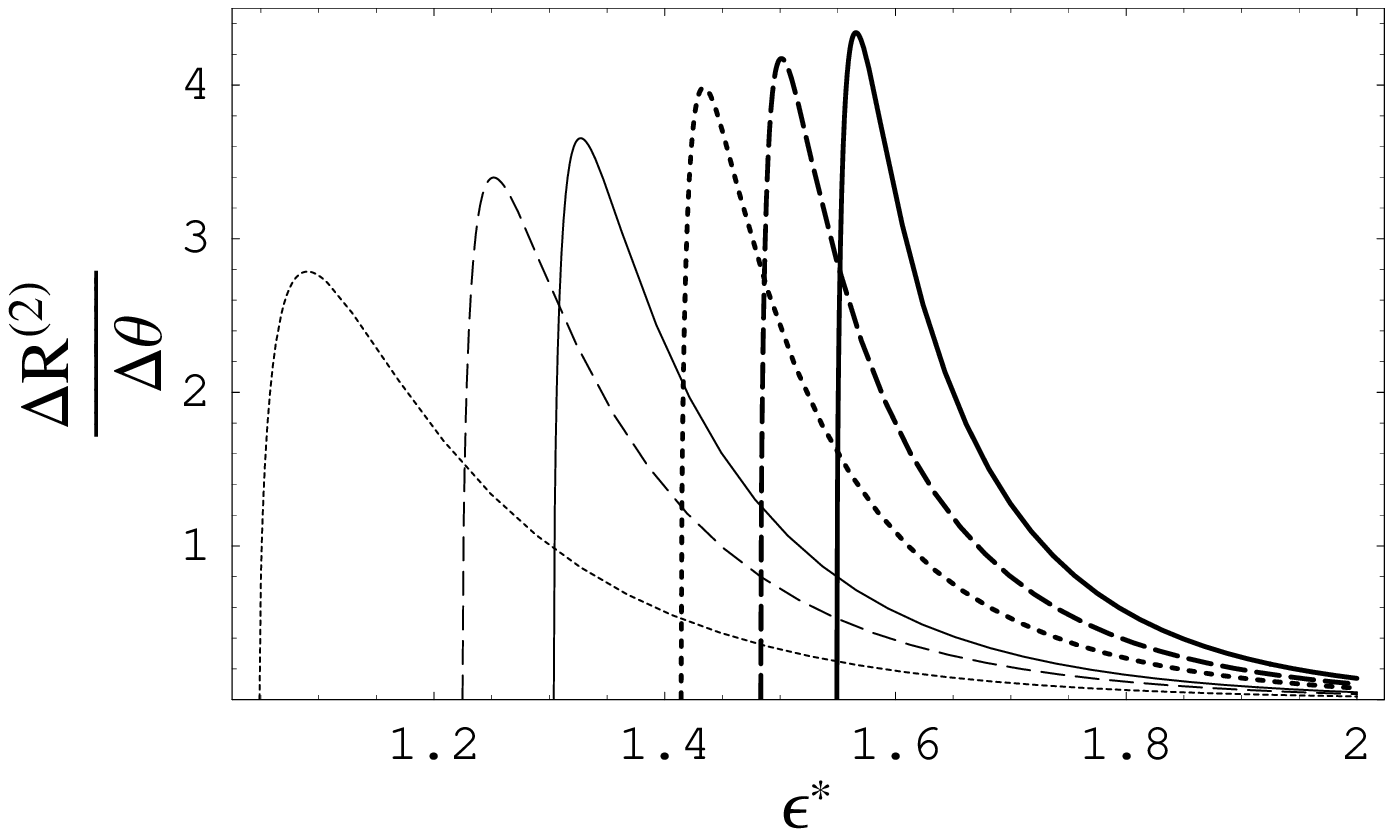}
\caption{Case $B\neq 0$ and $g = \Delta \theta f'$. $\Delta
R^{(s)} / \Delta \theta$ versus $\epsilon^*$ with $s=1$ (left
panel) and $s=2$ (right panel) for different values of the
magnetic field. Thick solid line: $b^* = 1.4$; thick dashed line:
$b^* = 1.2$; thick dotted line: $b^* = 1$; thin solid line: $b^* =
0.7$; thin dashed line: $b^* = 0.5$; thin dotted line: $b^* =
0.1$.}
\end{center}
\end{figure}


\begin{figure}[ht]
\begin{center}
\includegraphics[clip,width=0.48\textwidth]{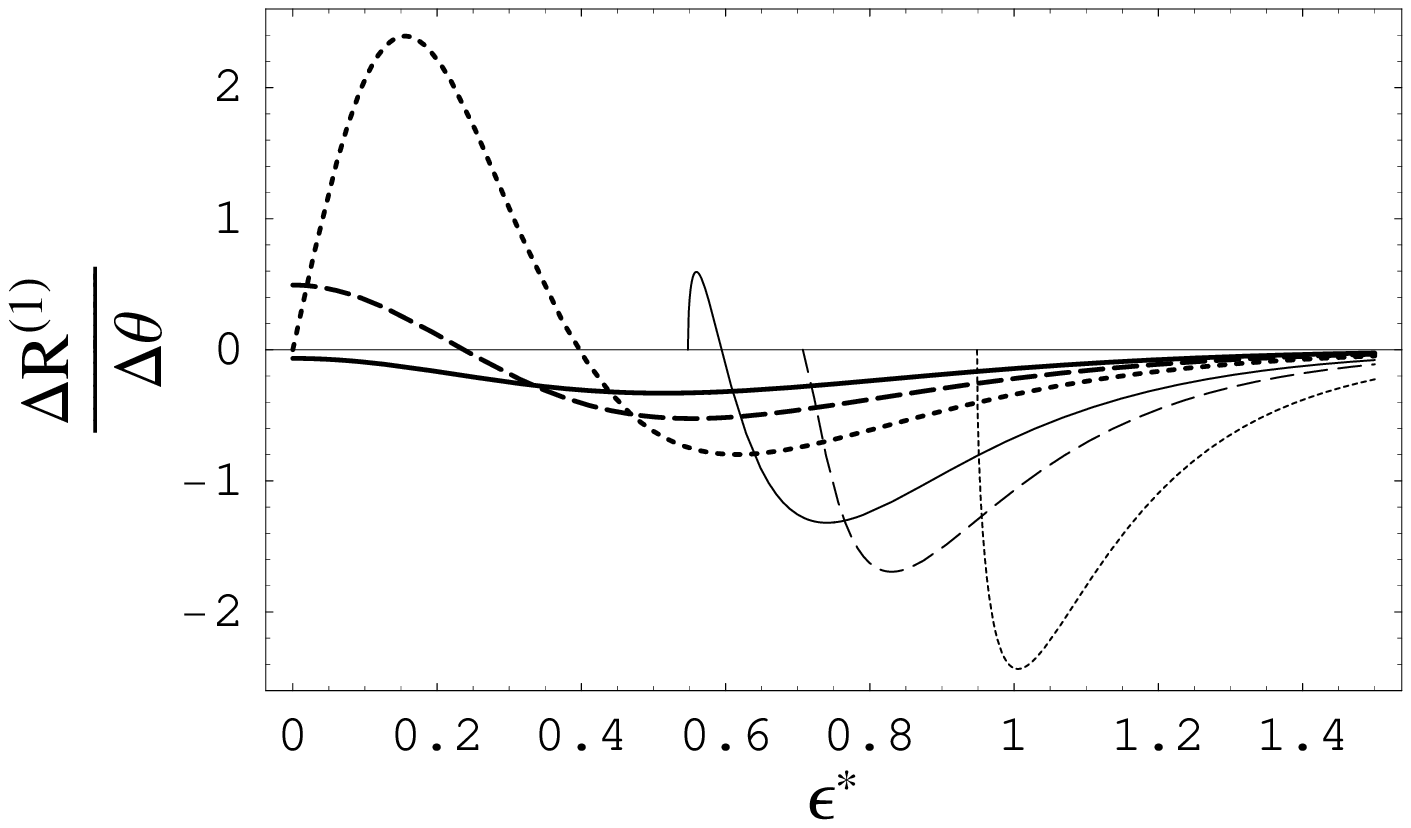}
\hskip 0.5truecm
\includegraphics[clip,width=0.48\textwidth]{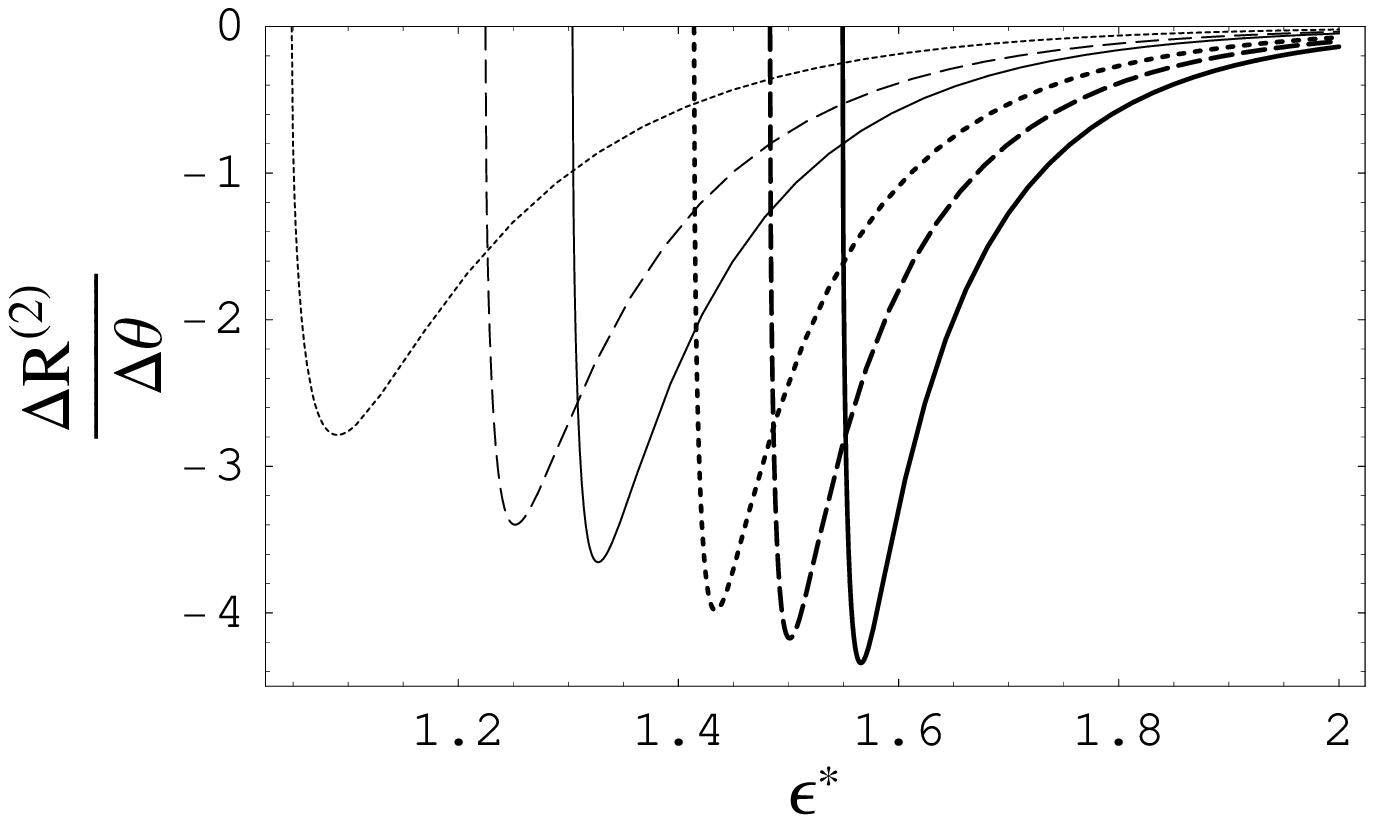}
\caption{Case $B\neq 0$ and $g = \Delta \theta f^2$. $\Delta
R^{(s)} / \Delta \theta$ versus $\epsilon^*$ with $s=1$ (left
panel) and $s=2$ (right panel). The values of $b^*$ are as in
Fig.~2.}
\end{center}
\end{figure}
%
%
Moreover we also have:
\begin{eqnarray}
\label{eq2.74a}
\delta_{V,s}^{\mbox{{\scriptsize inc}}} \!\!& = &\!\! \frac{\xi \,
b\, g_- }{2 \, \epsilon \, (\xi^2 - \alpha_s^2)} \,
 \: - \: (-1)^s \frac{
\xi}{\epsilon} \: \text{Re} \! \left[ \frac{\epsilon^2 -
(\alpha_s- \xi)^2} {2 \, \alpha_s^2} \: \, I^{(s)}_2 \,
\frac{\gamma_{-}(-\alpha_s, \alpha_s)}{\gamma_{+}(\alpha_s,
\alpha_s)} \: \right]
\\
\label{eq2.75a}
\delta_{V,s}^{\mbox{{\scriptsize refl}}} \!\!& = &\!\! \frac{\xi
\, b\, g_- }{2 \, \epsilon \, (\xi^2 - \alpha_s^2)} \,
 \: - \: (-1)^s \frac{
\xi}{\epsilon} \: \text{Re} \! \left[ \frac{\epsilon^2 -
(\alpha_s+ \xi)^2} {2 \, \alpha_s^2} \: \, I^{(s)}_2 \,
\frac{\gamma_{-}(-\alpha_s, -\alpha_s)}{\gamma_{+}(\alpha_s,
-\alpha_s)} \: \right]
\\
\label{eq2.76a}
\delta_{A,s}^{\mbox{{\scriptsize inc}}} \!\!& = &\!\! (-1)^s
\frac{2 \,  \epsilon \, \xi}{ \epsilon^2 + \xi^2 - \alpha_s^2}\:
\text{Re} \! \left[ \frac {\alpha_s - \xi} {\alpha_s}\, I^{(s)}_2
\,\frac{\gamma_{-}(-\alpha_s, \alpha_s)}{\gamma_{+}(\alpha_s,
\alpha_s)} \: \right]
\\
\label{eq2.77a}
\delta_{A,s}^{\mbox{{\scriptsize refl}}} \!\!& = &\!\! - (-1)^s
\frac{2 \, \epsilon \, \xi}{ \epsilon^2 + \xi^2 - \alpha_s^2}\:
\text{Re} \! \left[ \frac {\alpha_s + \xi} {\alpha_s}\, I^{(s)}_2
\,\frac{\gamma_{-}(-\alpha_s, -\alpha_s)}{\gamma_{+}(\alpha_s,
-\alpha_s)} \: \right]\, .
\end{eqnarray}
After some manipulations we obtain the explicit expression for
$\Delta R^{(s)}$:
\begin{equation}
\label{eq2.78} \Delta R^{(s)} = \frac{2R^{(0)}_s}{\varrho_s
-1/\varrho_s} \, (\delta_{V,s}^{\mbox{{\scriptsize inc}}} -
\delta_{V,s}^{\mbox{{\scriptsize refl}}} -
\delta_{A,s}^{\mbox{{\scriptsize inc}}} +
\delta_{A,s}^{\mbox{{\scriptsize refl}}}),
\end{equation}
where we have defined
\begin{equation}
\label{eq2.78bis}
R^{(0)}_s= \left| \frac {\gamma_+(\alpha_s, - \alpha_s)}
{\gamma_+(\alpha_s, \alpha_s)} \right|^2
\end{equation}
and
\begin{equation}
\label{eq2.78ter}
\varrho_s = (j^3_{V,s})_{(0)}^{\mbox{{\scriptsize
inc}}}/(j^3_{A,s})_{(0)}^{\mbox{{\scriptsize inc}}}\:.
\end{equation}
As for vanishing magnetic field, we consider the kink profile wall
Eq.~(\ref{kink}). In this case, the solutions of
Eq.~(\ref{eq2.19}), the explicit expressions for
$\gamma_{\pm}(\alpha_s, \alpha_s)$, and $R^{(0)}_s$ are the same
as in Eqs.~(\ref{eq2.80}), (\ref{eq2.81}), and (\ref{eq2.82})
respectively  with $\alpha$ replaced by $\alpha_s$.
\\
In Figs.~2 and 3 we display $\Delta R^{(s)}/ \Delta \theta$ as a
function of the energy $\epsilon^*$ with  magnetic field strength
$b^* = b {a^*}^2 = eB/m_0^2$. Again, we consider two different
functional forms for $g$: $g = \Delta \theta f'$ (Fig.~2) and $g =
\Delta \theta f^2$ (Fig.~3). For definitiveness  sake, we display
the curves for fixed wall thickness, $a^*=1$. Again we see that
the sign of $\Delta R$ depends  the functional form of $g$.
However, unlike the previous case $\Delta R$ for $s=1$ display
also a rather strong dependence on energy. Moreover, we see that
for high  energies of incident particles, $|\Delta R^{(s)}|$
rapidly decreases. These figures suggest that the presence of a
magnetic field can generate a reflection asymmetry between
particle with spin-up ($s=1$) and spin-down ($s=2$).


\renewcommand{\thesection}{\normalsize{\arabic{section}.}}
\section{\normalsize{Conclusions}}
\renewcommand{\thesection}{\arabic{section}}

In this paper we have discussed, within the so called non-local
baryogenesis where $CP$-violation and baryon number violation are
separated, the effects of a uniform magnetic field on the
scattering of fermions off $CP$-violating domain walls. The
$CP$-violation has been incorporated by considering a spatially
varying complex mass term $m(z)$.
We have calculated and discussed the asymmetry between the
reflection coefficients for right-handed and left-handed chiral
fermions, $\Delta R = R_{R \rightarrow L} - R_{L \rightarrow
R}\,$, which is of relevance in non local baryogenesis mechanisms.
In particular, we showed that a $z$ dependent phase in $m(z)$
implies that the reflection coefficients $R_{L \rightarrow R}$ and
$R_{R \rightarrow L}$ are different from each other leading to
$\Delta R \neq 0$. We have illustrated a general prescription to
solve the Dirac equation with a constant magnetic field treating
the $CP$-violating effects as a perturbation with respect to a
$CP$-conserving solutions.

The case of scattering of fermions off $CP$-violating bubble walls
in external magnetic field was studied in Ref. \cite{campanelli}.
\\
Our analysis reveals that there are some common points between
these two different cases. If we fix the value of the thickness of
the wall $a$ and the strength of the magnetic field $B$, $\Delta R
\rightarrow 0$ when the energy of the incident particles
approaches to infinity.
Moreover, the presence of a magnetic field generates a reflection
asymmetry between spin-up and spin-down particles. In particular,
the effect of the magnetic field is to shift the values of $\Delta
R^{(s)}$ (with respect to the case $B=0$) towards lower energies
in the case $s=1$, and higher energies in the case $s=2$.
Finally, the sign of $\Delta R$ varies with the thickness $a$ and
with the functional form of $g$ in both cases.
\\
Besides this common features, there are some important
differences.
\\
At fixed thickness of the wall and functional form of $g$ we have
a different behavior of $\Delta R$ as a function of the energy. In
fact, in the case of bubble walls the maximum value of $|\Delta
R^{(s)}|$ does not depend on $B$ and $s$ (for sufficiently high
values of energy). This is not the case for domain walls (see, for
example, Fig.~3).
Moreover, the physical situation studied in this paper is very
different from the case of bubble walls. In this last case we have
a phase transition proceeding by formation and expansion of
bubbles of new phase within the old ones in which the sphaleron
rate is suppressed. On the other hand, in the case of domain walls
the sphaleron mechanism is not exponentially suppressed in regions
where the vacuum expectation value of the Higgs field is small.
Therefore it is active only in the transition layer of the wall.
Finally, an interesting and peculiar phenomenon regarding kink
domain walls is the presence of localized states. An in depth
study of the domain wall-mediated electroweak baryogenesis for
generating a baryon asymmetry in the Universe must consider these
trapped fermions. We believe that this subject deserves further
investigations.

As is well known, the relevant global quantity in non-local defect
mediated electroweak baryogenesis is the flux of lepton number
radiated by the wall. In the case of bubble walls, the flux is
given by (see for instance Ref.~\cite{cohen,nelson}):
\begin{equation}
\label{eq3.1}
\Phi_L = \int \! \frac{d^3 k}{(2 \pi)^3} \; e^{-k/T} L({\textbf
k}) \cos \vartheta_R \, ,
\end{equation}
where $T$ is the temperature and $L({\textbf k}) = l |R({\textbf
k})|^2$;  $R({\textbf k})$ is the reflection amplitude for
particle of momentum ${\textbf k}$, $\vartheta_R$ is the angle of
reflection off the advancing wall, and $l$ is the lepton number.
In the case of an infinitely planar wall only the motion of
fermions perpendicular to the wall matter, so that we have:
\begin{equation}
\label{eq3.2}
\Phi_L (B)  \; \propto  \; \int \! dE \, e^{-E/T} \Delta R (E,B)
\; ,
\end{equation}
where $E$ is the energy of the scattered particle. Even though
non local baryogenesis mediated by planar kink domain walls has
been never discussed in the literature, it is conceivable that
Eq.~(\ref{eq3.2}) applies to planar kink domain walls. Actually,
the relevant quantity for the generation of the cosmological
baryon asymmetry turns out to be $\Phi_L^{tot}(B) +
\Phi_L^{tot}(-B)$, where
\begin{equation}
\label{eq3.3}
\Phi_L^{tot}(B) = \sum_{s=1,2} \Phi_L^{(s)}(B) \; \propto  \; \!
\sum_{s=1,2} \int \! dE \, e^{-E/T} \, \Delta R^{(s)}(E,B) \; ,
\end{equation}
with  $s$  the spin of the scattered particles. Indeed, it is
straightforward to verify that:
\begin{equation}
\label{eq3.4}
\Delta R^{(s)}(-B) = \Delta R^{(\overline{s})}(B),
\end{equation}
where $\overline{s} = 1$ if $s=2$, and $\overline{s} = 2$ if
$s=1$. So that, we have:
\begin{equation}
\label{eq3.5}
\Phi_L^{(s)}(B) + \Phi_L^{(s)}(-B) = \Phi_L^{(s)}(B) +
\Phi_L^{(\overline{s})}(B) = \Phi_L^{tot}(B),
\end{equation}
giving
\begin{equation}
\label{eq3.6}
\Phi_L^{tot}(B) + \Phi_L^{tot}(-B) = 2 \Phi_L^{tot}(B) \; .
\end{equation}
From Figures~2 and 3 we see that, in general, the quantity
$\Phi_L^{tot}(B)$ does not vanish. Thus, the total lepton number
flux radiated by a planar kink wall is different than zero.
However, it should be stressed that any realistic discussion of
non local baryogenesis mediated by planar kink domain walls needs
a careful treatment of the baryon number violating processes.

\newpage


\appendix{\normalsize{\bf {Appendix}}}
\section{\normalsize{Derivation of Equation~(\ref{eq2.52}) }}

In this Appendix we derive Eq.~(\ref{eq2.52}). To this end, we
show that, in general, the following expression:
\begin{equation}
\label{eqA1}
\frac{\gamma_{-}(\alpha_s,\alpha_s)}{\gamma_{+}(\alpha_s,\alpha_s)}
= \frac{\alpha_s + \xi}{\alpha_s - \xi}
\end{equation}
holds. Let us start by considering Eq.~(\ref{eq2.19}), that can be
written as:
\begin{equation}
\label{eqA3} D_{\mp} D_{\pm}  \phi^{(s)}_{\pm} = [\epsilon^2 -
(-1)^s b] \, \phi^{(s)}_{\pm} ,
\end{equation}
where we have introduced the operator $D_{\pm} = \mp D_x + \xi f$.
Multiplying Eq.~(\ref{eqA3}) by $D_{\pm}$, we get
\begin{equation}
\label{eqA5} (D_{\pm} D_{\mp}) [D_{\pm} \phi^{(s)}_{\pm}] =
[\epsilon^2 - (-1)^s b] \, [D_{\pm} \phi^{(s)}_{\pm}].
\end{equation}
The linearly independent solutions of Eq.~(\ref{eqA5}) are
$\phi^{(+\alpha_s)}_{\mp}$ and $\phi^{(-\alpha_s)}_{\mp}$. Hence,
comparing Eq.~(\ref{eqA3}) with Eq.~(\ref{eqA5})we see that
$D_{\pm} \phi^{(+\alpha_s)}_{\pm}$ can be written as a linear
combination of such solutions:
\begin{equation}
\label{eqA7}
D_{\pm} \phi^{(+\alpha_s)}_{\pm} = K^{(+\alpha_s)}_{\pm}
\phi^{(+\alpha_s)}_{\mp} + K^{(-\alpha_s)}_{\pm}
\phi^{(-\alpha_s)}_{\mp}.
\end{equation}
In order to determine the constants $K^{(\pm \alpha_s)}_{\pm}$, we
observe that in the limit $x \rightarrow +\infty$, we have $f(x)
\rightarrow 1$, $\phi^{(+\alpha_s)}_{\pm} \rightarrow e^{+\alpha_s
x}$, and $\phi^{(-\alpha_s)}_{\pm} \rightarrow e^{-\alpha_s x}$.
Therefore, we get $K^{(-\alpha_s)}_{\pm} = 0$ and
$K^{(+\alpha_s)}_{\pm} = \xi \mp \alpha_s$. In other terms, we
have:
\begin{equation}
\label{eqA10} ( \mp D_x + \xi f) \, \phi^{(+\alpha_s)}_{\pm} =
(\xi \mp \alpha_s) \, \phi^{(\alpha_s)}_{\mp}.
\end{equation}
Now, considering the asymptotic behavior  for $x \rightarrow -
\infty$
\begin{equation}
\label{eqA11} \lim_{x \rightarrow -\infty}
\phi^{(+\alpha_s)}_{\pm}(x) = \gamma_{\pm}(\alpha_s,\alpha_s) \,
e^{\alpha_s x} + \gamma_{\pm}(\alpha_s, -\alpha_s) \, e^{-\alpha_s
x},
\end{equation}
and taking into account the lower sign in Eq.~(\ref{eqA10}), we
have
\begin{eqnarray}
\label{eqA12} (\alpha_s - \xi) \, [\gamma_{-}(\alpha_s, \alpha_s)
\, e^{\alpha_s x} + \gamma_{-}(\alpha_s, -\beta_s) \, e^{-\alpha_s
x}] && \nonumber \\
&&
\!\!\!\!\!\!\!\!\!\!\!\!\!\!\!\!\!\!\!\!\!\!\!\!\!\!\!\!\!\!\!\!\!\!\!
\!\!\!\!\!\!\!\!\!\!\!\!\!\!\!\!\!\!\!\!\!\!\!\!\!\!\!\!\!\! =
(\alpha_s + \xi) \, [\gamma_{+}(\alpha_s, \alpha_s) \, e^{\alpha_s
x} + \gamma_{+}(\alpha_s, -\alpha_s) \, e^{-\alpha_s x}].
\end{eqnarray}
Equation~(\ref{eqA1}) follows immediately from the above
relations.
\\
We can now obtain Eq.~(\ref{eq2.52}). Inserting
Eqs.~(\ref{eq2.47}) and (\ref{eq2.50}) into Eq.~(\ref{eq2.16}) we
get:
\begin{eqnarray}
\label{eqA13} [\Psi_s(x,\tau)]^{\mbox{{\scriptsize inc}}} \!\!& =
&\!\! A^{(+)}_s \gamma_+(\alpha_s, \alpha_s) \, e^{-i\epsilon \tau
+ \alpha_s x}
\nonumber \\
\!\!& \times &\!\! \left \{ \left[ \epsilon - (-1)^s \xi g_{-} +
\frac{(-1)^s \xi (\alpha_s - \xi)}{2\alpha_s } \left(
\frac{\gamma_{-}(-\alpha_s, \alpha_s)}{\gamma_{+}(\alpha_s,
\alpha_s)} + c^{(+)}_{-} \frac{\gamma_-(\alpha_s,
\alpha_s)}{\gamma_+(\alpha_s, \alpha_s)} \right) \! I^{(s)}_2
\right] \! u^s_{-} \right.
\nonumber \\
\!\!& + &\!\! \left. \left[ -\alpha_s -\xi + \frac{(-1)^s \xi
\epsilon (\alpha_s - \xi)}{2\alpha_s} \left(
\frac{\gamma_{-}(-\alpha_s, \alpha_s)}{\gamma_{+}(\alpha_s,
\alpha_s)} + c^{(+)}_{-} \frac{\gamma_{-}(\alpha_s, \alpha_s)}
{\gamma_+(\alpha_s, \alpha_s)} \right) \! I^{(s)}_2 \right] \!
u^s_+ \right \}. \nonumber
\\
&& {\phantom {.}}
\end{eqnarray}
The first term containing $c^{(+)}_{-}$ in Eq.~(\ref{eqA13}) is
modified taking into account Eq.~(\ref{eqA1})
\begin{equation}
\label{eqA14} \Theta \equiv \frac{(-1)^s \xi (\alpha_s -
\xi)}{2\alpha_s} \, \frac{\gamma_{-}(\alpha_s, \alpha_s)}
{\gamma_+(\alpha_s, \alpha_s)} \, I^{(s)}_2 c^{(+)}_{-} =
\frac{(-1)^s \xi (\alpha_s + \xi)}{2\alpha_s} \, I^{(s)}_s
c^{(+)}_{-}.
\end{equation}
Following the same arguments presented in Appendix of
Ref.~\cite{funakubo}, we argue that $(\alpha_s + \xi) \, I^{(s)}_2
c^{(+)}_{-}/\alpha_s$ can be substituted with  $g_{-}$. Therefore,
we have $\Theta = (-1)^s \xi g_{-}/2$. Moreover, the terms
containing $c^{(+)}_{-}$ in Eq.~(\ref{eqA13}) can be handled in
the same manner. In conclusion we have:
\begin{equation}
\label{eqA17} \frac {(-1)^s \xi \epsilon}{2\alpha_s} \, \frac
{\gamma_{-}(\alpha_s, \alpha_s)}{\gamma_{+}(\alpha_s, \alpha_s)}
\, I^{(s)}_2 c^{(+)}_{-}  =  \frac {(-1)^s \xi
\epsilon}{2(\alpha_s - \xi)} \: g_{-} \, .
\end{equation}
Taking into account these last results, we easily recover the wave
function Eq.~(\ref{eq2.52}).

\section{\normalsize{Derivation of Equation~(\ref{eq2.68a})}}
In this appendix we derive Eqs.~(\ref{eq2.68a}) and
(\ref{eq2.69a}). Let us  consider the vectorial current
$j^{\mu}_V= \bar{\Psi} \, \gamma^{\mu} \, \Psi$. After  taking
into account Eq.(\ref{eq2.52}) and Eq.(\ref{eq2.53}), it is
straightforward to obtain the incident and reflected vectorial
currents:
\begin{eqnarray}
\label{eqB.1} (j^3_{V,s})^{\text {inc}} \!\!& =\!\! &  2 \,\,
\epsilon \,\, |A^{(+)}_s \, \, \gamma_+(\alpha_s, \alpha_s)|^2
\,\, \text{Im} \left[\, - \alpha_s - \xi + \frac {\xi \,  b \, g_-
} {2 \, \epsilon \, (\alpha_s - \xi)} \, \, + \right. \nonumber
\\
& & \left.  - \frac {\xi  (-1)^s} { 2 \, \epsilon \,  \alpha_s}
\,\, {I^{(s)}_2}^* \, (\alpha_s + \xi)^2 \, \frac
{\gamma_-(\alpha_s, -\alpha_s)} { \gamma_+(-\alpha_s, -\alpha_s)}
+ \frac {\xi \, \epsilon \, (-1)^s} { 2 \, \alpha_s} \, I^{(s)}_2
\,  \,  \frac {\gamma_-(-\alpha_s, \alpha_s)} { \gamma_+(\alpha_s,
\alpha_s)} \, \right] \nonumber
\\
& & \phantom {xx}
\end{eqnarray}
\begin{eqnarray}
\label{eqB.2} (j^3_{V,s})^{\text {refl}} \!\!& =\!\! &  -2 \,\,
\epsilon \,\, |A^{(+)}_s \, \, \gamma_+(\alpha_s, -\alpha_s)\, |^2
\,\, \text{Im} \left[\, \alpha_s - \xi - \frac {\xi \,  b \, g_- }
{2 \, \epsilon \, (\alpha_s + \xi)} \, \, + \right. \nonumber
\\
& & \left.  - \frac {\xi  (-1)^s} { 2 \, \epsilon \,  \alpha_s}
\,\, {I^{(s)}_2}^* \, ( -\alpha_s + \xi)^2 \, \frac
{\gamma_-(\alpha_s, \alpha_s)} { \gamma_+(-\alpha_s, \alpha_s)} +
\frac {\xi \, \epsilon \, (-1)^s} { 2 \, \alpha_s} \, I^{(s)}_2 \,
 \,  \frac {\gamma_-(-\alpha_s, -\alpha_s)} { \gamma_+(\alpha_s,
-\alpha_s)} \, \right]\, . \nonumber
\\
& & \phantom {xx}
\end{eqnarray}
The axial currents $j^{\mu}_A= \bar{\Psi} \, \gamma^{\mu} \,
\gamma_5 \Psi$ can be handled in the same way. We have:
\begin{equation}
\label{eqB.3} (j^3_{A,s})^{\text{inc}}=(-1)^{s+1} \,\,  \epsilon^2
\, |A^{(+)}_s \, \, \gamma_+(\alpha_s, \alpha_s)\, |^2  \left\{ 2-
\frac {(-1)^s \, b} {\epsilon^2} + (-1)^s \frac {2 \, \xi}
{\epsilon} \, \text{Re} \left[ I^{(s)}_2 \, \frac {\alpha_s - \xi}
{\alpha_s} \, \frac {\gamma_- (-\alpha_s, \alpha_s)}
{\gamma_+(\alpha_s, \alpha_s)} \right] \right\}
\end{equation}
\begin{equation}
\label{eqB.4} (j^3_{A,s})^{\text{refl}}=(-1)^{s+1} \,\,
\epsilon^2 \, |A^{(+)}_s \, \, \gamma_+(\alpha_s, -\alpha_s)\, |^2
\left\{ 2- \frac {(-1)^s \, b} {\epsilon^2} - (-1)^s \frac {2 \,
\xi} {\epsilon} \, \text{Re} \left[ I^{(s)}_2 \, \frac {\alpha_s +
\xi} {\alpha_s} \,  \frac {\gamma_- (-\alpha_s, -\alpha_s)}
{\gamma_+(\alpha_s, -\alpha_s)} \right] \right\}.
\end{equation}
Our goal is to show that:
\begin{equation}
\label{eqB.5} (j^3_{V,s})^{\mbox{{\scriptsize inc}}}  =
(j^3_{V,s})_{(0)}^{\mbox{{\scriptsize inc}}} \, (1 +
\delta_{V,s}^{\mbox{{\scriptsize inc}}})
\end{equation}
with $(j_V^3)^{inc}_{(0)}$ and $\delta_V^{inc}$ given by Eq.(\ref
{eq2.70a}) and Eq.(\ref {eq2.74a}) respectively, since the other
cases can be obtained similarly. We observe that Eq.(\ref {eqB.1})
can be written  as Eq.(\ref {eqB.5}) once we put:
\begin{equation}
\label{eqB.6} (j^3_{V,s})^{inc}_{(0)} = 2 \, \epsilon \,
|A^{+}_{s}|^2 \, |\gamma_+(\alpha_s,\alpha_s)|^2 \, \, \text{Im}(-
\alpha_s) = - 2\,
 i \, \epsilon \,  \alpha_s \, |A^{+}_{s}|^2 \,
|\gamma_+(\alpha_s,\alpha_s)|^2 \; ,
\end{equation}
and
\begin{eqnarray}
\label{eqB.8} \delta_{V,s}^{\text{inc}}&=& - \frac {1}
{|\alpha_s|} \, \text{Im} \left[ \frac { \xi \, b \, g_-} {2 \,
\epsilon \,  (\alpha_s - \xi)} + (-1)^s \frac {\xi \, \epsilon }
{2 \, \alpha_s} z- (-1)^s \frac {\xi \, (\alpha_s+ \xi)^2} {2 \,
\epsilon \,  \alpha_s}\,  z^* \right] \nonumber
\\
&=&- \frac {1} {|\alpha_s|} \, \text{Im} \left\{ \frac {(-1)^s \,
\xi \,  \epsilon} {2 \,  \alpha_s} \left[ (-1)^s \, \frac {b \,
g_- \, \alpha_s } {\epsilon^2\,  (\alpha_s - \xi)} + z - \frac
{(\alpha_s + \xi)^2} {\epsilon^2}\,  z^* \right] \right\}\, ,
\end{eqnarray}
where
\begin{equation}
\label{eqB.7} z \equiv I_2^{(s)} \, \frac {\gamma_-(-\alpha_s,
\alpha_s)} {\gamma_+(\alpha_s,\alpha_s)}\, .
\end{equation}
Indeed, Eq.~(\ref{eqB.6}) coincides with Eq.~(\ref{eq2.70a}). To
obtain Eq.~(\ref{eq2.75a}) we note that, given a real number $a$,
we have $\text{Im}(i \, a \, z)= a \, \text{Re}$. So that, we can
write Eq.~(\ref{eqB.8}) as:
\begin{equation}
\label{eqB.9} \delta_{V,s}^{\text{inc}} = \frac {(-1)^s \, \xi\,
\epsilon} {2 \, |\alpha_s|^2} \, \text{Re}\left[\frac {(-1)^s \,
\alpha_s \, b \, g_-} {\epsilon^2 \, (\alpha_s - \xi)}+ z -
\left(\frac {\alpha_s + \xi} {\epsilon}  \right)^2 z^* \right].
\end{equation}
Observing that $\text{Re}
[\alpha_s/(\alpha_s-\xi)]=|\alpha_s/(\alpha_s - \xi)|^2$, we have
\begin{equation}
\label{eqB.10} \delta_{V,s}^{\text{inc}}= \frac {\xi \, b \, g_-}
{2 \, \epsilon \, |\alpha_s - \xi|^2} + \frac {(-1)^s \xi \,
\epsilon} {2 \, |\alpha_s|^2} \text{Re} \left[z -\left(\frac
{\alpha_s + \xi} {\epsilon} \right)^2 z^* \right] \; .
\end{equation}
On the other hand
\begin{equation}
\label{eqB.11} \text{Re}\left[\left(\frac {\alpha_s + \xi}
{\epsilon}\right)^2 \, z^* \right] = \frac {\xi^2- |\alpha_s|^2}
{\epsilon^2}\, \text{Re} [z] + \frac {2 \, |\alpha_s| \, \xi}
{\epsilon^2} \, \text{Im}[z] \; ,
\end{equation}
whereupon
\begin{equation}
\label{eqB.12} \delta_{V,s}^{\text{inc}} = \frac {\xi \, b \, g_-}
{2 \, \epsilon \, |\alpha_s - \xi|^2} + \frac {(-1)^s \xi \,
\epsilon} {2 \, |\alpha_s|^2}\left\{\frac {\epsilon^2 - \xi^2 +
|\alpha_s|^2} {\epsilon^2 }\, \text{Re[z]} - \frac {2 |\alpha_s|
\, \xi} {\epsilon^2} \, \text{Im}[z] \right\} \; .
\end{equation}
Finally, after taking into account that:
\begin{equation}
\label{eqB.13} \frac {\epsilon^2 - \xi + |\alpha_s|^2} {2\,
|\alpha_s| \, \xi} \, \text{Re} [z] + \text{Re} [iz]=
\text{Re}\left[ \frac {\epsilon^2 - (\alpha_s-\xi)^2} {2 \,
|\alpha_s|\, \xi} \, z\right] \; ,
\end{equation}
we recast Eq.~(\ref{eqB.12}) into:
\begin{equation}
\label{eqB.14} \delta_{V,s}^{\mbox{{\scriptsize inc}}} =
\frac{\xi \, b\, g_- }{2 \, \epsilon \, (\xi^2 - \alpha_s^2)} \,
 \: - \: (-1)^s \frac{
\xi}{\epsilon} \: \text{Re} \! \left[ \frac{\epsilon^2 -
(\alpha_s- \xi)^2} {2 \, \alpha_s^2} \: \, I^{(s)}_2 \,
\frac{\gamma_{-}(-\alpha_s, \alpha_s)}{\gamma_{+}(\alpha_s,
\alpha_s)} \: \right] \;,
\end{equation}
which indeed agrees with Eq.~(\ref{eq2.74a}).

\newpage


\renewcommand{\thesection}{\normalsize{\arabic{section}.}}

\end{document}